\documentclass[runningheads]{llncs}
\usepackage{graphicx}
\usepackage{enumitem}  
\usepackage{tabularx}
\usepackage[linesnumbered,ruled,vlined]{algorithm2e}
\usepackage{wrapfig,lipsum,booktabs}
\usepackage{multicol}
\usepackage{adjustbox}

\begin{document}

\title{Improving IT Support by Enhancing Incident Management Process with Multi-modal Analysis}
%
%
\author{Atri Mandal\inst{1} \and Shivali Agarwal\inst{1} \and Nikhil Malhotra\inst{2}  \and Giriprasad Sridhara\inst{1} \and Anupama Ray\inst{1} \and Daivik Swarup\inst{1}}
\authorrunning{A. Mandal et al.}
%
\institute{IBM Research AI, Bengaluru, India \and
IBM Global Technology Services, Bengaluru, India
\email{\{atri.mandal,shivaaga,nikhimal,girisrid,anupamar,dvenkata\}@in.ibm.com}}
\maketitle              
\begin{abstract}
\footnote{\textbf{This paper has been accepted for presentation in International Conference on Service Oriented Computing (ICSOC) 2019 – to be held in Toulouse, France; 28-31 October, 2019. This is an author copy. The respective Copyrights are with Springer}}
IT support services industry is going through a major transformation with AI becoming commonplace. There has been a lot of effort in the direction of automation at every human touchpoint in the IT support processes. Incident management is one such process which has been a beacon process for AI based automation. The vision is to automate the process from the time an incident/ticket arrives till it is resolved and closed. While text is the primary mode of communicating the incidents, there has been a growing trend of using alternate modalities like image to communicate the problem.
A large fraction of IT support tickets today contain attached image data in the form of screenshots, log messages, invoices and so on. These attachments help in better explanation of the problem which aids in faster resolution. 
Anybody who aspires to provide AI based IT support, it is essential to build systems which can handle multi-modal content. \par
In this paper we present how incident management in IT support domain can be made much more effective using multi-modal analysis. 
The information extracted from different modalities are correlated to enrich the information in the ticket and used for better ticket routing and resolution. 
We evaluate our system using about 25000 real tickets containing attachments from selected problem areas. Our results demonstrate significant improvements in both routing and resolution with the use of multi-modal ticket analysis compared to only text based analysis. 

\keywords{Service delivery \and Incident Management \and  Multimodal Analysis \and Image Understanding \and Automated routing and resolution}
\end{abstract}
%
%
%
\section{Introduction}
\label{sec:intro}
Incident management process in modern IT service delivery is undergoing a massive transformation with an ever increasing focus on automation of tasks that require human cognizance. Two such key tasks are that of {\it ticket assignment} and {\it resolution} as they require considerable amount of manual labour. There are quite a few recent instances in the service industry where assignment/resolution has been automated using analysis of structured and unstructured text content.  All these systems generally work for text content only. However, a lot of these tickets have attachments of pictures, screenshots, logs etc. 
which not only help in giving a visual representation of the problem but also provide necessary context information. For example, 
an end user needing troubleshooting assistance for a software application (e.g. out of memory issue) will take a screenshot capturing the error message (and error code, if any) and the running application(s) along with CPU/memory usage statistics. Resolution of tickets without considering such important details may not only result in an unsatisfactory resolution, but can also mislead or confuse the user, leading to poor customer experience and multiple escalations. Also in a lot of cases textual information may be completely absent from the ticket and the troubleshooting agent has to infer the problem only from the attachments.
In all these scenarios, it is important to address the fundamental problem of understanding the screenshot  images, extract the relevant information and generate problem descriptions which can then be utilized in the automation pipeline. \par
\begin{figure}
\centering
{\includegraphics[height=3cm,width=.47\columnwidth]{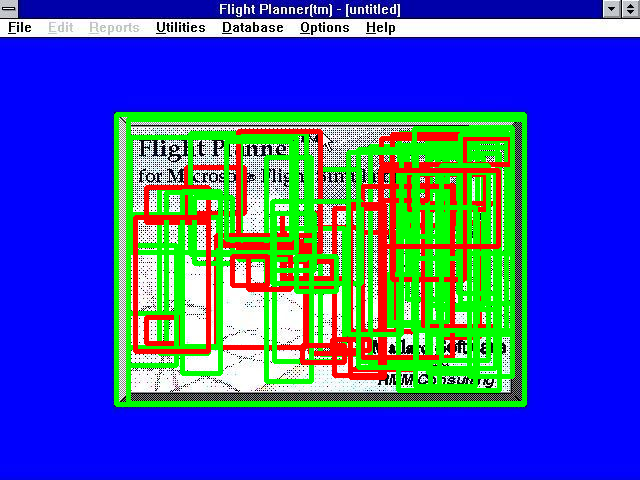}}
{\includegraphics[height=3cm,width=.47\columnwidth]{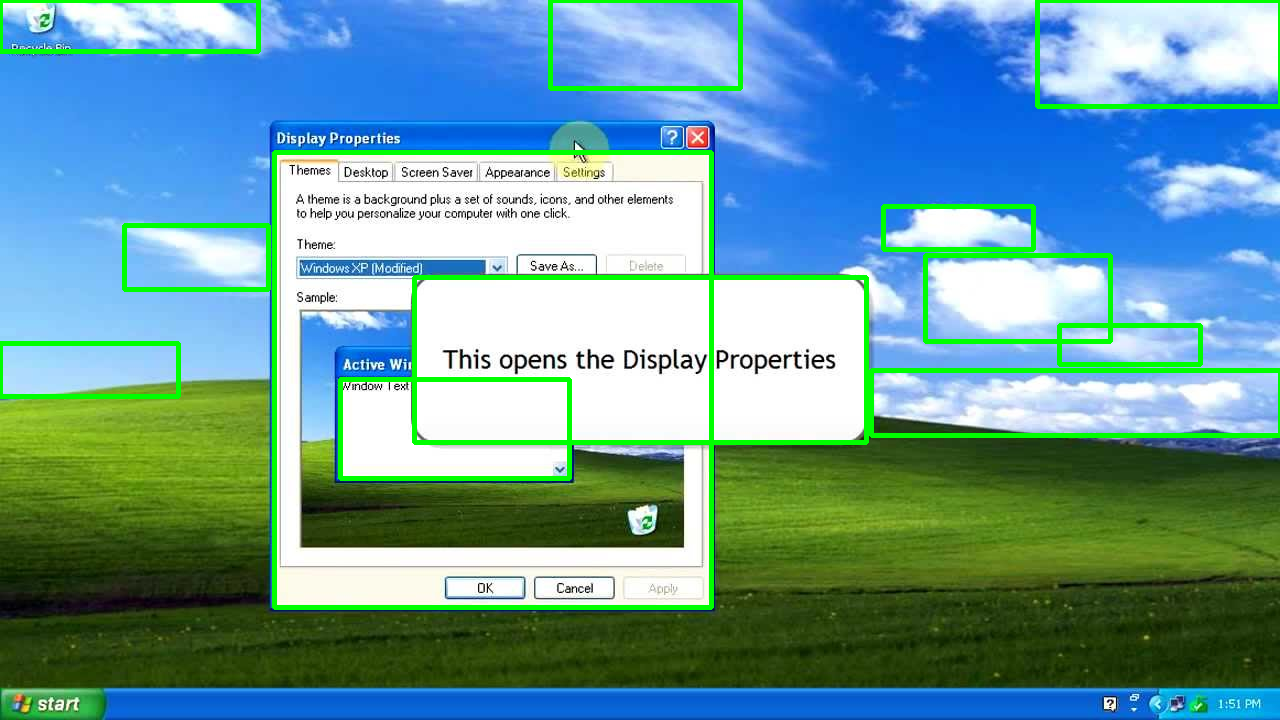}}
\caption{(a) Canny-edge detects spurious boxes (b) Contour detection detects objects in background}
\label{fig:contour_detection_err}
\end{figure}

There are a quite a few challenges in extracting information from screenshot images and using them in a proper way to arrive at a resolution. Some of these challenges are: 
i) Lack of labeled training data with images/videos annotated for the boxes with important information or labels in the form of actual content of images (text groundtruth). To the best of our knowledge there is no such annotated dataset available for IT support domain with labeled images. Thus, deep learning models,  which require a lot of training data cannot be trained on this domain with multimodal data.
ii) Presence of overlapping windows often occludes the text content which might be relevant for better assignment or resolution. Thus conventional image processing algorithms like  contour detection (\cite{Maire:2009:CDI:1925546}) or canny edge detection (\cite{Canny:1986:CAE:11274.11275}) do not work well only by themselves and fail to understand the internal structure or content in the windows (as shown in Figure \ref{fig:contour_detection_err}).
iii) 
To obtain the embedded text in the image we can use Optical Character Recognition (\cite{Mori:1999:OCR:319799}).
However the image may have a lot of noisy text which are not related to the problem (e.g. icon labels, menu items, code, console commands etc.) and so the complete text obtained from OCR may not be useful. 
iv) The correlation between ticket text and textual content extracted from the image is also challenging as domain knowledge plays a very important part in this correlation and content understanding.


In this paper, we discuss an end-to-end system which can analyze image content in tickets, understand the nature of the problem indicated in the image and automatically suggest a resolution. In this paper we focus on a specific type of attachment, viz. screenshots, as this is the most common type of attachment, requiring human supervision, found in IT support tickets. The key contributions of our paper are described below:
\begin{enumerate}[label=(\roman*)]
  \item A high-precision hybrid object detection engine which uses a combination of traditional image processing algorithms as well as deep learning based image classification. The main purpose of the detection engine is to identify if an application window (e.g. error message box, terminal, explorer window etc.) is present and if so, the type of the application window. 
  \item A ticket enrichment module which uses OCR and NLP based techniques to extract relevant pieces of information from the application window(s) detected in the image and uses this extra information to enrich the ticket data for better classification.
  \item A scalable routing and resolution recommendation framework, having an intelligent decision making mechanism based on its confidence on multiple predicted fields. 
\end{enumerate}

Using our system we were able to demonstrate significant improvements in both ticket assignment and ticket resolution accuracy compared to only text based analysis. The automation achieved by our system can result in an estimated saving of 200000 man hours per annum for a helpdesk account receiving 100000 tickets a month. 

The rest of the paper is organized as follows. Section \ref{sec:related_work} discusses some of the related work in the area. Section \ref{sec:system} gives an overview of the system architecture used. In Section \ref{sec:evaluation}  we present our experimental results while we conclude in Section \ref{sec:conclusion}.


\section{Related Work}
\label{sec:related_work}
Incident management process has been discussed in literature with a focus on ticket categorization/problem determination, ticket dispatch/resolver group prediction, resolver group formation and resolution recommendation. 
Many systems proposed in the past provide a solution for automated problem determination and resolution e.g. ~\cite{DasguptaIcsoc14}, \cite{7877279} talk about auto-remediation by first categorizing the ticket into a problem category and then recommending a solution for the problem category identified. They have used text based classification. The system in \cite{7505624} proposes resolution recommendation for event tickets using an approach which utilizes both
the event and resolution information in historical tickets via
topic-level feature extraction. 
The work in \cite{Zhou:2017:SST:3097983.3098190} also proposes a solution for automated ticket resolution using a deep neural network ranking model trained on problem text and resolution summary of historical tickets. ReACT system~\cite{AggarwalSCC16} performs an involved natural language processing to help create resolution sequences for ticket categories in a semi-automated way. However all the above mentioned systems analyze only the text part of the ticket. Analysis of images have not been dealt with in these systems.\par
In another body of work, there is a focus on the ticket dispatch and resolver group aspects. SmartDispatch
\cite{AgarwalKdd} provides a solution for automated ticket dispatch using Support Vector Machines and discriminative keyword approach. Historical data on agents and their current workloads is used for ticket dispatch in \cite{Botezatu2015}. More recently, the system in \cite{DBLP:conf/icsoc/MandalMARS18} uses a combination of rule engine and ensemble classifier to achieve very high accuracy in resolver group prediction. 
However none of these works analyze the screenshots and attachments that often contain vital information.  \par
There are also systems which have looked solely at the problem of mining information from images. However most of the literature deals with mining, extracting or summarizing information from natural images which cannot be used directly due to the challenges stated in section~\ref{sec:intro}. There is very little work done in the past which focuses on extracting information from technical screenshots. Anand et al. \cite{AnandSampat_Stanford} is one such paper. However, it only mines the screenshots to broadly classify the application and does not deal with occlusion and text correlation. Senthil et al. (\cite{DBLP:journals/corr/abs-1711-02012}) proposes a Question-Answering (QA) system for ticket resolution where they look at image screenshots containing error. However the system has looked at specific types of errors (SAP) and rely solely on OCR to retrieve errors from images. These systems also do not handle occlusion and text inferences.\\
We have not come across any work that performs multi-modal(text+image) analysis on ticket data addressing the challenges of occlusion, text enrichment and correlation like we have done in this paper. Our proposed approach is generic enough to be applied to chatbots and QA systems. 

\section{Multi-modal Analysis in Incident Management}
\label{sec:system}
\begin{figure} 
{\includegraphics[width=.8\textwidth]{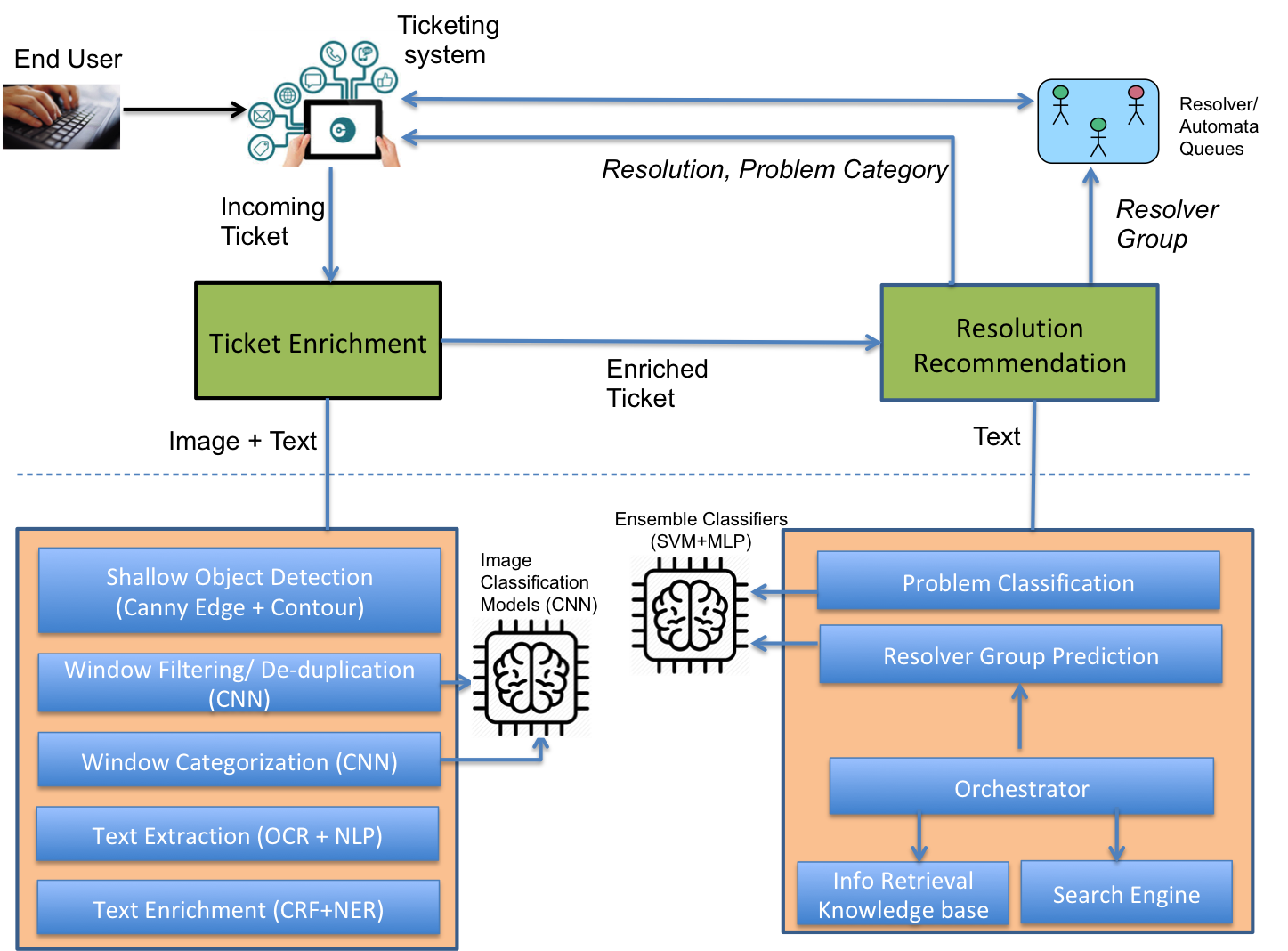}}
\centering
\caption{System architecture}
\label{fig:run_time_arch}
\end{figure}
The traditional lifecycle of incident management has undergone massive changes in recent times due to the infusion of agent assist capabilities.  
The motivation is to i) automate ticket assignment and resolution with high accuracy whenever possible and ii) reduce the time taken to resolve in case of manual resolution. These objectives are primarily achieved through two functional modules viz. Ticket Enrichment and Resolution Recommendation.  
The incident management lifecycle with agent assist capabilities is depicted in Figure~\ref{fig:run_time_arch}. 
The ticket enrichment module uses models trained on historical data to enrich ticket data with knowledge inferred from the ticket data. The resolution recommendation module leverages the enriched ticket information to predict the most accurate resolution with high confidence. Once the ticket is augmented with inferred knowledge on resolution and problem category, it is stored in the system and the agents can leverage it for speedy resolution. 

We now explain how the ticket enrichment is done using multi-modal analysis, that is, combined analysis of text and image present in the ticket. We also explain the proposed multi-step process for resolution recommendation which can choose the source of resolution based on the confidence on its own knowledge.
\subsection{Ticket Enrichment}
Often users are unaware of the exact problem or do not know what all details might be important for solving the problem and end up not specifying relevant information. For example, in a lot of IT support tickets the name of the operating system, application, version and other important contextual information are omitted. Without these information it may be difficult to drill down to the exact problem category and resolver group. Thus, we augment the text data with context information and insights obtained from the image data to create a better ticket which helps in improving the prediction of  resolver group and problem category leading to faster ticket resolution. The different stages of the ticket enrichment pipeline are described in detail below. \par
\textbf{Image Understanding}:
The image understanding part analyzes the attachment image and extracts artifacts which are used for understanding the image properties. The most important information in a screenshot is usually contained within one of the application windows. Therefore, one of the key functionalities of our system is to detect an application window. We also classify the detected window based on its type e.g. browser/IDE, console or dialog/message box. We now describe the image understanding steps below.\par
\textbf{i. Shallow object detection}: 
\label{section:shallow_object_detection}
The objective of this stage is to detect the precise coordinates of the window objects present in the screenshots. We experimented with two well documented computer vision techniques for object detection viz. \textit{Contour Detection} and \textit{Canny Edge detection} as described below. \par

\textit{Contour Detection}:
\label{sec:contour}
Contour detection \cite{Maire:2009:CDI:1925546} is used to detect objects with both linear and non-linear contours. Before applying contour detection the input image is transformed using i) Gaussian blur and ii) binary conversion. This method suffers from two major drawbacks. Firstly, this method not only detects rectangular boxes but also objects with irregular shapes which may be present in the picture as illustrated in Figure \ref{fig:contour_detection_err}. To solve this problem we use a shape detector to detect relevant objects of rectangular shape. But this still does not exclude the possibility of detecting rectangular non-window objects, so we often end up with false positives. Secondly, detection of window fails when the colors of the background and the object to be detected are roughly similar resulting in both objects being converted to the same color during binarization. 

\textit{Canny Edge Detection}: 
We also use an alternate method for window detection viz. Canny Edge detector using Hough lines \cite{Duda:1972:UHT:361237.361242}.
This technique can detect all horizontal and vertical lines in a picture and as such can be used to detect regular geometrical shapes e.g. triangles or rectangles. 
Before applying canny edge detector we convert the image to grayscale. The detected lines are clustered based on their coordinates to detect rectangular shapes. 
However canny edge detection fails when windows do not have clear demarcating lines. Also in some cases canny edge detection ends up mining spurious boxes as shown in Figure \ref{fig:contour_detection_err}. 

To increase the accuracy of shallow detection we use an ensemble of both techniques. However, even with the ensemble the precision is low as none of the shallow detection methods look at the internal structure of the window. To reduce false positives and improve precision we use a filtering step as described below.

\textbf{ii. Window filtering/deduplication}:
We use different filtering technologies to remove spurious and duplicate windows detected in the previous step. We first use a size based filter to remove all windows which are smaller than a threshold. This removes GUI artifacts like radio buttons, alert/minimize/cancel icons etc. We then use a CNN based binary image classifier on the filtered boxes to classify whether the box is an actual application window or not. We use a CNN based binary classification model, which is trained using screen shots of end-user problems downloaded from the net and also on synthetically generated windows. For feature extraction we use ResNet50 model \cite{He2016DeepRL} pre-trained with ImageNet weights. We prefer using ResNet50 architecture over VGG19 \cite{Simonyan15} as it uses skip connections to handle the problem of vanishing gradients. For classification we added two fully connected layers. The classifier layer was fine tuned during training and feature extractor layer was frozen. Our model is able to indicate presence/absence of application window with an accuracy of about 95\%. \\
Finally we apply a de-duplication step to remove duplicate windows. Since both shallow detection techniques are applied independently and in parallel there is the possibility of detecting the same window twice. Duplicate windows can be detected based on the coordinates of the enclosing rectangle and calculating the area of overlap using IOU metric. 

\textbf{iii. Deep learning based Window Categorization}\\ 
In this stage of the pipeline we try to categorize the detected windows as well as identify certain window properties for deeper understanding of the image. Previously there has been work on identifying application name and other properties from text part of the image (\cite{DBLP:journals/corr/abs-1711-02012}). However in the case where one or more application windows are overlapped the text in the background window will be occluded and may not be useful for extraction. We take the help of deep learning to try and identify these properties upfront. \par
We make use of two separate classifiers for this step. The first classifier is used for classification of windows into specific categories to identify the application type. We support only a few selected applications as of now but our classifier can be easily extended to support more applications. The second classifier is used to determine the OS (Windows, Linux, Mac).  We used 1 CNN block having a convolution layer followed by ReLU activation, max-pooling and batch normalization for feature extraction followed by two fully-connected (FC) layers for classification. \par 

\textbf{Text Extraction from Images}:
Once the window categorization and segmentation phase is over, text is detected and recognized using Tesseract OCR \cite{Mori:1999:OCR:319799} 
from the detected application windows. Since we are dealing with screenshots, the resolution of the image was not an issue.  
Due to challenges of overlapping windows/boxes or errors in window detection, the text extraction is not accurate. We use two different types of post-processing on the recognized text. Firstly we use a dictionary based post-processing step (using edit-distance) to correct spellings errors for application names or title boxes. For longer text (e.g. dialog box, console logs etc) we use a word-level language model trained on a very large data of logs and error messages from stack-overflow. This language model not only helps us improve word error rates but also predicts words in occluded windows. We observed in our results that if the text is occluded by a line, we were able to recover it but if the box suffers from a higher overlap the text does not get fully recovered even by the language model.

\textbf{Ticket Text Enrichment}:
In this step, we enrich the ticket text with information extracted from the image. However we cannot directly use all the text extracted from the image for ticket enrichment. In order to extract key terms  and entities  we use a Conditional Random Fields (CRF) based Named Entity Recognition (NER) system \cite{gupta-etal-2018-semantic} on both ticket text as well as all text extracted from images. This extractive system gives us terms such as name of operating system(OS), application/product name, components being mentioned, version numbers, error codes, error messages and other entities such as symptoms or important mentions from log screenshots. For OS name, application name and components, domain specific dictionaries are used and for version and error codes we use regular expression based extraction. For the other attributes such as symptom, activity, action and advise we use deep parsing and understanding \cite{gupta-etal-2018-semantic}.
We then correlate these entities with the information obtained from the image to retain only the most relevant parts of the image information. The resulting text is then inserted into the ticket using slot-based templates for ticket completion. The slot templates can differ based on the resolver group. The examples below illustrate the technique of slot-filling for ticket enrichment. The enriched parts of the email are enclosed within square braces and the slot names are mentioned in angular braces along with the corresponding values.\\
\begin{scriptsize} \textbf{Example1:}  {\fontfamily{qcr}\selectfont "Dear sir, My postpaid mobile \textbf{[$<$mobile-no$>$ = xxx3224]} having relationship number \textbf{[$<$customer-no$>$ = xxx]}, billing plan \textbf{[$<$billing-plan$>$ = infinityxxx]} has been overcharged with international roaming services \textbf{[$<$pack-details$>$ = international roaming XXX nrc]} for the billing period \textbf{[$<$period$>$ = 08-jan-2019 to 07-feb-2019]} which was not activated by me. You can clearly find the same in the screen shot of bill details sent. Please refund me the overcharged charges asap. Regards, xxx xxx mobile ---  xxx3224"} \\
\textbf{Example2:} {\fontfamily{qcr}\selectfont "I am getting an error \textbf{[$<$errmsg$>$ = An error occurred during the installation of assembly component HRESULT: 0x800736FD]} with error code \textbf{[$<$ errcode$>$ = Error 1935]}, while installing \textbf{[$<$appname$>$ = Crystal Reports Runtime Engine]} for .Net on \textbf{[$<$os$>$ = Windows]} \textbf{[$<$osver$>$ = 10]}. Please see attached screenshot"} \end{scriptsize} \\


\subsection{Resolution Recommendation System}
For resolution of tickets we use a recommendation engine which reads the tickets enhanced with information from the ticket enrichment module, understands the user intent and uses it to suggest the most relevant resolver group and resolution(s). The recommendation system is trained using a corpus of historical tickets T which is divided into two parts viz. T\textsubscript{H} (\textit{short head}) and T\textsubscript{L} (\textit{long tail}). T\textsubscript{H} contains the most frequently occurring problem categories having a well known resolution and typically accounts for 75-80\% of the tickets. T\textsubscript{L} constitutes the rarely occurring problem categories for which a well curated resolution may or may not be present in our training corpus. The division of tickets is done according to the following equation: 
\begin{equation}
    T = T_{H} + T_{L}
\end{equation}
\begin{equation}
    T_{H} = \bigcup_{p_{i} \in P_{H}} T_{p_{i}} 
\end{equation}
where $P_{H}$ is the set of problem categories in short head and $T_{p_{i}}$ is the set of tickets belonging to the problem category $p_{i}$. It's important to note that problem category may be a composite field in the ticketing system. In this case we concatenate the constituent sub-field labels to obtain the unique problem category for training. \\
To select $P_{H}$ we plot a histogram of frequencies for problem category and select the ones which are above a configured threshold. We also do some post processing to filter out those categories which do not have well defined resolutions. 
We use separate strategies for resolving the \textit{short head} and the \textit{long tail} tickets as described below. 

\textbf{Ticket classification}:
The objective of ticket classification is to predict the resolver group and the problem category. We train an ensemble classifier using only the data in $T_{H}$. This reduces noise in training data and also eliminates class imbalance problem \cite{2018arXiv180802636M}. For the ensemble classifier we use simple classification models viz. Linear SVM (ovr) and MLP (feed forward neural nets) for easy deployability and retraining \cite{2018arXiv180802636M}. We plotted the accuracy and coverage of the selected classifiers against different confidence thresholds and selected the optimal threshold value to ensure that both classifiers in the ensemble operate at least at human level efficiency \cite{2018arXiv180802636M}.

\textbf{Ticket resolution}:
To obtain a resolution at runtime we first use our ensemble classifiers to predict the resolver group and problem category. If both these fields are predicted with high confidence at runtime it means that the problem category belongs to the short head. In this case we return a resolution directly using a simple database lookup. If the confidence score for the resolver-group or the problem category is low then we resort to our long tail approach which queries the knowledge corpus ingested through an information retrieval infrastructure (e.g. Watson Discovery).  
We observe that while we have a resolution available for most frequent short head queries, we may not have them for infrequent or unseen queries. To handle this case, we use a web search and combine the retrieved resolutions with web search results using the enriched ticket description as query. We re-rank the combined results and present the top $N$ results to the user. For this, we use a federated search algorithm. \\
We build a resource representation for ticket content and web resources by sampling tickets and related web search documents respectively. For each, we compute the unigram distribution of terms. Using this unigram language model, we compute the relevance score for tickets as well as for resources from web. 
We then use the CORI result merging algorithm\cite{CORI} to merge the results using the relevance scores to obtain the final ranked list as shown in Equation \ref{eq:cori}, where d is the normalized score given by the search engine and c is the relevance score computed by the language model.\\
\begin{equation}
    \label{eq:cori}
    result\_score = \frac{d + 0.4 \times c \times d}{1.4}
\end{equation}
The different steps in the resolution process is orchestrated by the \textbf{orchestrator} which is the key computational module of the recommendation system. The complete ticket resolution process is explained in detail in Algorithm \ref{algo:end-to-end}.
\newsavebox{\tempbox}
\savebox{\tempbox}{
\SetInd{0.25em}{0.2em}
\begin{minipage}[c]{0.45\textwidth}%
\begin{scriptsize}
\begin{algorithm*}[H]
\label{algo:end-to-end}
\SetKwInOut{Input}{Input}\SetKwInOut{Output}{Output}
\Input{Enriched ticket text}
\Output{result = [ resolv\_grp, prob\_category, resolution ]}
\SetKwFunction{FMain}{ticket-assignment-resolution}
\SetKwProg{Fn}{Function}{:}{}
\BlankLine
\Fn{\FMain{ Enriched-Email-Text}}{
\BlankLine
final\_result = [None, [], []]
\BlankLine
classification = InvokeCombinedClassifier( Enriched-Email-Text )
\BlankLine
\If{(classification.conf\_resolv\_grp $>$ CONF\_RESOLV\_CUTOFF) \textbf{and}  (classification.conf\_prob\_category $>$ CONF\_PROB\_CUTOFF)} {
    \tcc{short head - directly lookup resolution result}
    resolution = lookup(resolutionDB, classification.prob\_category)
    \BlankLine
    final\_result = [classification.resolv\_grp, classification.prob\_category, resolution]
}
\Else  {
    \tcc{invoke long tail strategy}
    filter\_fields = []
    \BlankLine
    \If { classification.conf\_resolv\_grp $>$ CONF\_RESOLV\_CUTOFF } {
        filter\_fields += resolv\_grp
        \BlankLine
        final\_resolv\_grp = classification.resolv\_grp
    }
    \Else {
        \tcc {Assign ticket to manual queue}
         final\_resolv\_grp = None
    }
    \For{each subfield in PROBLEM\_CATEGORY\_FIELD} {
        result = InvokeProblemClassifier(subfield, Enriched-Email-Text)
        \BlankLine
        \If{ result.conf\_subfield $>$ CONF\_SUBFIELD\_CUTOFF } {
            filter\_fields += subfield
        }
    }
    \tcc{invoke Information Retrieval and web search in parallel and combine/rerank results}
    searchRes = InvokeSearch( filter=filter\_fields, text=Enriched-Email-Text)
    \BlankLine
    webSearchRes = InvokeWebSearch( Enriched-Email-Text)
    \BlankLine
    fedSearchRes = InvokeFederatedSearch( Enriched-Email-text, searchRes, webSearchRes)
    \BlankLine
    final\_result = [final\_resolv\_grp, fedSearchRes.prob\_category, fedSearchRes.resolution]
  }
return final\_result
}
\end{algorithm*}%
\end{scriptsize}
\end{minipage}
}

\begin{algorithm}
\caption{Ticket Resolution Algorithm}
\clipbox{0pt {\depth} 0pt {\baselineskip}}{\usebox{\tempbox}}\hfill
\raisebox{\depth}{\clipbox{0pt 1ex 0pt {\height}}{\usebox{\tempbox}}}
\end{algorithm}

\begin{figure}
   \setlength\abovecaptionskip{0.5\baselineskip}
    \setlength\belowcaptionskip{0pt}
   {\includegraphics[height=5cm,width=0.9\textwidth]{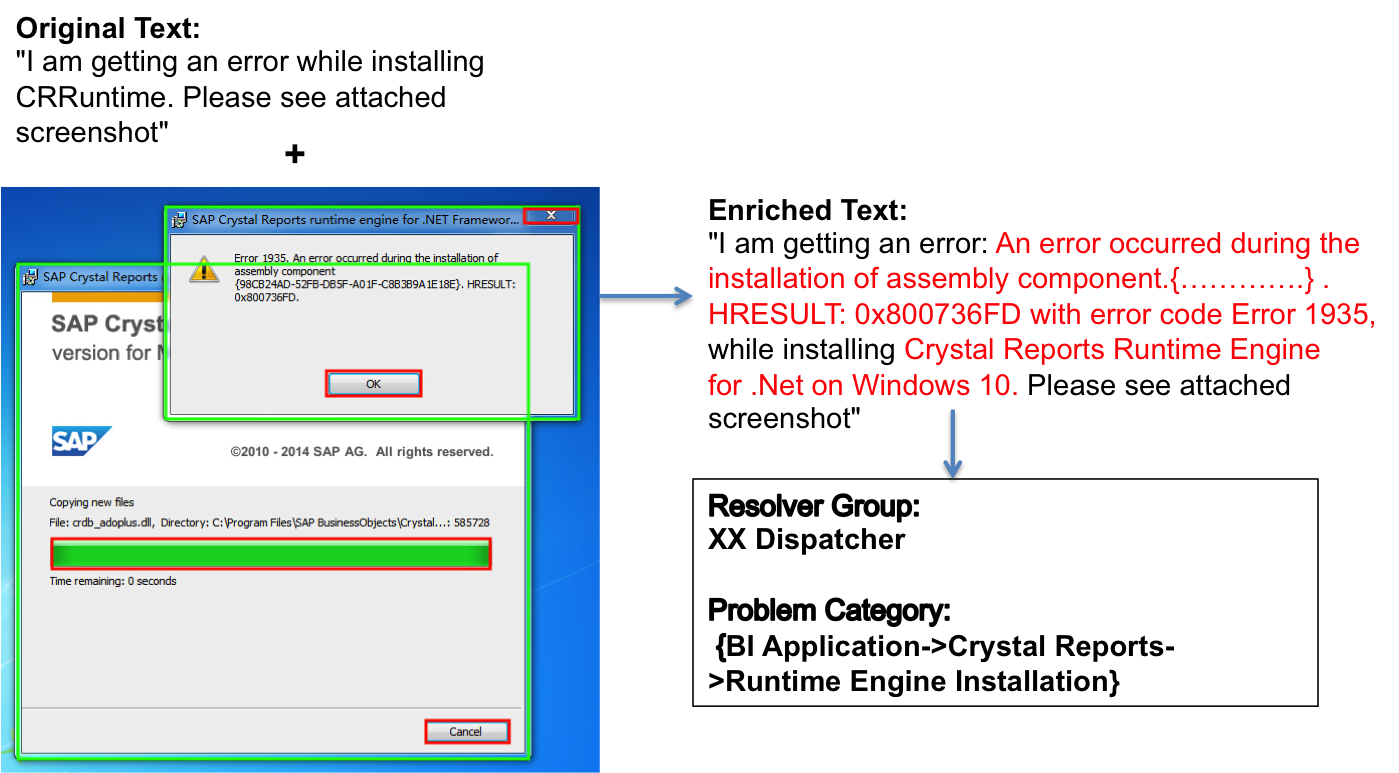}} 
   \centering
   \caption{Illustrative example}
   \label{fig:motivating_example}
\end{figure}


\section{Dataset details and Experiment Setup}
\subsection{IT Support ticket data}
Our evaluation is based on a ticketing dataset having a corpus of 712320 support tickets from 428 resolver groups and spanning 3728 distinct problem categories as shown in Table \ref{tab:dataset}. Out of this corpus 159344 tickets (approx. 22.37\%) had attachments.  However for this paper we limited our scope to a small subset of this dataset mainly because the image understanding part of our system currently does not handle all possible type of applications. To select our experimental dataset we chose 10 resolver groups with the maximum amount of screenshot attachments. Out of these resolver groups we chose the most frequently occurring 33 problem categories for our short head training dataset. The remaining tickets accounting for 219 problem categories constitutes the long tail. The total number of multimodal tickets in our curated dataset is 25000. 

\subsection{Image Data}
\textbf{Collection}:
The image data for our training is mainly obtained from the attachments in the ticketing dataset. However to increase the volume for training as well as to get more variety in training data we also scraped relevant images from the web (Google Images). We used a search filter to download images for only selected applications. Apart from this we also generated synthetic screenshot images using a python library (pySimpleGUI). Using this library we can easily control image parameters like size and coordinates of the generated window, text content, size and count of radio buttons etc.\\ 
\textbf{Augmentation}:
To enhance the size of our training set we used both offline and online image augmentation. We perform the following transformations on each image to generate new images offline, viz. changing brightness and contrast levels, conversion to grayscale and resizing. Apart from these transformations we also use Keras augmentation API for further augmentation of the images during the training process. \\
\textbf{Annotation}:
Annotation of image data is a laborious process as it involves manual annotation of bounding boxes for windows as well as embedded image text. For both these types of annotation we used automation. \\
For bounding box annotation we used shallow object detection technique described in \ref{section:shallow_object_detection}. This method of annotation works on most images. However whenever images contain windows with high degree of overlap and confusing images in the background the annotation may not be entirely correct. In these cases we do a manual inspection and annotation.\\ 
For getting ground truth data on image text we primarily use synthetically generated screenshots with pre-defined text content. In this case both the window and the text are generated by our script and no manual annotation is necessary. For real screenshots, we first perform OCR on the image and then manually correct the extracted text to generate groundtruth.
\begin{scriptsize}
\begin{table}
\centering
\small
\caption{Dataset Details}
  \label{tab:dataset}
  {%
  \begin{tabularx}{0.95\textwidth}{|X|X|X|X|}
    \hline
    \textbf{} & \textbf{Total Tickets} & \textbf{Problem Categories} & \textbf{Multimodal Tickets} \\
    \hline
    \textbf{Overall} & 712230 & 3728 & 159344 \\
    \hline
    \textbf{Selected } & 42882 & 252 & 25000 \\
    \hline
    \end{tabularx}
  }
\smallskip
\centering
\small
\caption{Accuracy of shallow object detection }
  \label{tab:opencvobj}
  {%
  \begin{tabularx}{0.95\textwidth}{|X|X|X|X|}
    \hline
    \textbf{Method} & \textbf{1-Window(P,R)} & \textbf{2-windows(P,R)} & \textbf{3-windows(P,R)}\\
    \hline
    \textbf{Contour} & 70\%,76\% & 62\%,78\% & 57\%,68\% \\
    \hline
    \textbf{Canny edge} & 43\%,82\% & 53\%,80\% & 48\%,64\% \\
    \hline
    \textbf{Ensemble+Filter} & 90\%,89\% & 90\%,86\% & 92\%,72\% \\
    \hline
    \end{tabularx}
  }
\smallskip 
\centering
\small
\caption{Accuracy of Image classification}
  \label{tab:windowcat}
  {%
  \begin{tabularx}{0.95\textwidth}{|X|X|X|X|}
    \hline
    \textbf{Method} & \textbf{Window Filtering} & \textbf{Operating System} & \textbf{Application Category} \\
    \hline
    \textbf{VGG19} & 92.3\% & 91.5\% & 85.7\%\\
    \hline
    \textbf{ResNet50 } & 94.9\% & 94.1\% & 90.8\%\\
    \hline
    \end{tabularx}
 }
\smallskip 
\centering
\small
\caption{Dataset Accuracy}
  \label{tab:acc}
  {%
  \begin{tabularx}{0.95\textwidth}{|X|X|X|}
    \hline
    \textbf{} & \textbf{Text Only} & \textbf{Multimodal} \\
    \hline
    \textbf{Assignment(acc/cov)} & 86.1\%,89.3\% & 88.6\%,96.5\% \\
    \hline
    \textbf{Resolution} & 74.7\% & 82.4\% \\
    \hline
    \end{tabularx}
}
\end{table}
\end{scriptsize}
\\
\textbf{Experimental setup}:
For our deep learning based experiments we used a NVIDIA Tesla K80 GPU cluster with 4 CUDA-enabled nodes. For the remaining experiments we used a IBM softlayer VM having 256G RAM, 56 CPU cores and 100G HDD.

\section{Evaluation}
\label{sec:evaluation}

Figure \ref{fig:motivating_example} illustrates the working of our pipeline with a real example. The bounding boxes detected by our system are indicated in green while those which are filtered out after detection are indicated in red. Interestingly shallow object detection detects the green sliding status bar which is eventually filtered by our deep learning based window filtering technique. Our system is not only able to detect the error message box correctly but also the box in background which has relevant context information. We highlight some of the important context information picked up by our system. Combining the information in the detected windows the system is able to suggest the most relevant troubleshooting page for the error. Evaluation of the different functional stages of our multimodal analysis pipeline is presented below.\\
\textbf{Detection of windows}: To detect window objects we first experimented with DL based object detection. However we observed that training the object detection algorithm using traditional image datasets like MSCOCO (\cite{10.1007/978-3-319-10602-1_48}) and ImageNet (\cite{imagenet_cvpr09}) does not result in high accuracy. One of the reasons is that deep learning based methods usually need a large number of training samples and it is difficult to obtain such a large corpus to train.  Also the objects in these datasets correspond to natural images with widely different features than those available in screenshots.\\ 
As far as shallow object detection is concerned both canny edge detection and contour detection suffer from the problem of high recall/low precision. This is because both these methods detect objects without understanding the internal structure resulting in false positives.
However a combination of the  techniques improves both precision and recall significantly as shown in Table \ref{tab:opencvobj}. \\
\textbf{Image classification}: For our DL based image classification models we experimented with various hyper-parameters like learning rate(LR), filters, filter size, number of neurons etc. We found LR to be the topmost contributor in accuracy. We ran LR range test and plotted the accuracy versus LR, noting the LR value when the accuracy starts to increase and when the accuracy becomes ragged \cite{learning-rates-smith-2017}. Our results in Table \ref{tab:windowcat} indicate very high accuracy (more than 90\%) for image classification with ResNet50. Since the images have large inter-class variance and small intra-class variance we also experimented with shallow CNNs and VGG19. However, with limited amount of training data ResNet50 (with pre-trained weights) proved to be a better choice than its shallow counterparts. The result means our system can identify the application type and OS accurately in more than 90\% of the cases even when window text is occluded. \\
\textbf{Text extraction}: We evaluated the correctness of our text extraction technique using mainly synthetic images to avoid manual annotation. Synthetic images with pre-defined text content were generated using OpenCV python libraries and the generated text was compared with that obtained from OCR. We used two different OCR techniques for our evaluation viz. Watson Visual Recognition and Tesseract, out of which Tesseract performed better. Our OCR technique was observed to have more than 95\% accuracy (character level). However we also manually corroborated the results with real data for a few images. \\ 
\textbf{Routing/Resolution}
To evaluate the accuracy of resolution we look at the classification results for resolver group and problem category. As routing is a key step in the resolution of the ticket we have to ensure that routing of the ticket is improved by our multimodal analysis technique.\\
Also, the most important step in obtaining the resolution strategy is to understand the correct problem category of the ticket as in most cases, the  problem category has a one-to-one mapping with the resolution strategy. Even if that is not the case, identifying the correct problem category is a key step in automated resolution as it narrows down the scope of the search. As such we estimate the accuracy of resolution with the accuracy achieved in predicting the problem category in both the short head and long tail cases.
The results are shown in Table \ref{tab:acc}. For our dataset the \textit{problem category} is a composite field constituting three sub-fields. 
We consider the identified problem category to be accurate if and only if all the three sub-fields were identified correctly. Using this metric we achieved an overall accuracy of 82.4\% with multimodal, an improvement of about 8\% over text based analysis.  In fact, for some problem categories belonging to the long tail the observed improvement was more than 50\% proving that multimodal analysis is helpful in automated resolution of tickets. Prediction accuracy of resolver group also improves by about 2.5\% but more importantly the automation coverage increases by more than 7\% as more tickets are predicted with higher confidence. Considering that these improvements are over and above an already deployed system (using text-based analysis), the numbers are significant.

\subsection{Impact to Incident Management process}
We calculate the impact to the incident management process based on two aspects viz. Routing and Resolution. For our dataset the incoming rate of tickets is approximately 100,000 per month. We assume that a human agent takes about 3 min to read and assign each ticket and 10 min to actually resolve the ticket. On the basis of the above assumptions the net savings for an account can be calculated as: 
\begin{equation}
   S_{assign} = N \times T_{cov} \times 3
\end{equation}
\begin{equation}
   S_{resolve} = N \times R_{cov} \times 10
\end{equation}
where $N$ is the total number of tickets per annum, $T_{cov}$ is the coverage for automated routing, $R_{cov}$ is the coverage for automated resolution, $S_{assign}$ is the net savings from routing and $S_{resolve}$ is the net savings from resolution. This gives a total saving of about 194,000 man hours per annum assuming $T_{cov}=90\%$ and 
$R_{cov}=80\%$

\section{Conclusion and Future Work}
\label{sec:conclusion}
In this paper we have presented an end-to-end system which can analyze image content in ticket attachments, enrich ticket text and automatically suggest a resolution. As of now we have limited our scope to analyzing only images with screenshots. In reality there may be many different types of attachments with varying properties and user intent. Some of these images may require deep understanding of the layout or semantic structure of the image. For example, sales related support issues may require processing of invoices containing tables, bar charts etc having a specific layout. Without understanding the layout we cannot analyze the document for troubleshooting. 
In the future we will look at advanced computer vision techniques to understand and analyze such types of attachments. 
 
{\small
\bibliographystyle{splncs04}
\bibliography{multimodal_references}
}

\end{document}